\documentclass[11pt]{article}
\usepackage{graphicx}
\usepackage{cite}

\newcommand{\BABARPubYear}    {06}

\newcommand{\BABARConfNumber} {034}
\newcommand{\SLACPubNumber} {12025}

\def\numBB{347 million}

\def\GeV{\;\mbox{GeV}}

\def\GeVcc{\;\mbox{GeV}/c^2}

\def\MeVcc{\;\mbox{MeV}/c^2}

\def\thd  {\theta_D}
\def\de   {\Delta E}

\def\mes  {M_{\mbox{\scriptsize ES}} }
\def\bkg  {B \to K^{*}\gamma}

\def\bkgneut  {B^0 \to K^{*0}\gamma}
\def\bkpg  {\bu \to K^{*+}\gamma}

\def\brpg {B^+ \to \rho^+\gamma}
\def\brzg {B^0 \to \rho^0\gamma}
\def\bomg {B^0 \to \omega\gamma}

\def\BFrp{1.06^{+0.35}_{-0.31}\pm 0.09} 
\def\BFrz{0.77^{+0.21}_{-0.19}\pm 0.07} 
\def\BFom{0.39^{+0.24}_{-0.20}\pm 0.03} 
\def\BFomUL{0.84}                       
\def\BFav{1.01\pm 0.21\pm 0.08} 
\def\effrp{11.6}                        
\def\effrz{14.5}                        
\def\effom{8.1}                         
\def\BrhoBKst{0.024\pm0.005}                
\def\BrhozBKstz{0.038^{+0.011}_{-0.010}}   
\def\VtdVtsval{0.171^{+0.018+0.017}_{-0.021-0.014}}   
\def\VtdVtsvalLabeled{0.171^{+0.018}_{-0.021}\mathrm{(exp.)}^{+0.017}_{-0.014}\mathrm{(theor.)}}   
\def\VtdVtsRzVal{0.216^{+0.029+0.021}_{-0.031-0.018}} 

\def\bsg    {b\to s\gamma}

\def\rhoz {\rho^0}
\def\rhop {\rho^+}

\def\bdg    {\ensuremath {b \to d \gamma}}
\def\bkg    {\ensuremath {\B \to \Kstar \gamma}}
\def\bkog    {\ensuremath {\Bz \to \Kstarz \gamma}}

\def\bkpg    {\ensuremath {\Bp \to \Kstarp \gamma}}

\def\Kz    {\ensuremath{K^{0}}\xspace}

\def\de        {\ensuremath {\Delta E}}

\def\avbr      {\ensuremath{\overline{\BR}[B\rightarrow(\rho/\omega)\gamma]}}
\def\VtdVts    {\ensuremath{|V_{td}/V_{ts}|}}

\RequirePackage{xspace}

\usepackage{relsize}
\def\babar{\mbox{\slshape B\kern-0.1em{\smaller A}\kern-0.1em
    B\kern-0.1em{\smaller A\kern-0.2em R}}}

\def\en         {\ensuremath{e^-}\xspace}   
\def\ep         {\ensuremath{e^+}\xspace}
 
\def\epem       {\ensuremath{e^+e^-}\xspace}

\def\g     {\ensuremath{\gamma}\xspace}

\def\s     {\ensuremath{s}\xspace}

\def\b     {\ensuremath{b}\xspace}

\def\piz   {\ensuremath{\pi^0}\xspace}

\def\pip   {\ensuremath{\pi^+}\xspace}
\def\pim   {\ensuremath{\pi^-}\xspace}

\def\pipm  {\ensuremath{\pi^\pm}\xspace}

\def\Kbar  {\kern 0.2em\overline{\kern -0.2em K}{}\xspace}

\def\Kz    {\ensuremath{K^0}\xspace}
\def\Kzb   {\ensuremath{\Kbar^0}\xspace}
\def\KzKzb {\ensuremath{\Kz \kern -0.16em \Kzb}\xspace}
\def\Kp    {\ensuremath{K^+}\xspace}
\def\Km    {\ensuremath{K^-}\xspace}
\def\Kpm   {\ensuremath{K^\pm}\xspace}

\def\KpKm  {\ensuremath{\Kp \kern -0.16em \Km}\xspace}
\def\KS    {\ensuremath{K^0_{\scriptscriptstyle S}}\xspace} 
 
\def\Kstarz  {\ensuremath{K^{*0}}\xspace}

\def\Kstar   {\ensuremath{K^*}\xspace}

\def\Kstarp  {\ensuremath{K^{*+}}\xspace}

\def\Dbar    {\kern 0.2em\overline{\kern -0.2em D}{}\xspace}

\def\Dz      {\ensuremath{D^0}\xspace}
\def\Dzb     {\ensuremath{\Dbar^0}\xspace}
\def\DzDzb   {\ensuremath{\Dz {\kern -0.16em \Dzb}}\xspace}
\def\Dp      {\ensuremath{D^+}\xspace}
\def\Dm      {\ensuremath{D^-}\xspace}

\def\DpDm    {\ensuremath{\Dp {\kern -0.16em \Dm}}\xspace}

\def\B       {\ensuremath{B}\xspace}
\def\Bbar    {\kern 0.18em\overline{\kern -0.18em B}{}\xspace}

\def\BB      {\ensuremath{B\Bbar}\xspace} 
\def\Bz      {\ensuremath{B^0}\xspace}
\def\Bzb     {\ensuremath{\Bbar^0}\xspace}
\def\BzBzb   {\ensuremath{\Bz {\kern -0.16em \Bzb}}\xspace}
\def\Bu      {\ensuremath{B^+}\xspace}
\def\Bub     {\ensuremath{B^-}\xspace}
\def\Bp      {\ensuremath{\Bu}\xspace}

\def\BpBm    {\ensuremath{\Bu {\kern -0.16em \Bub}}\xspace}

\def\BorBbar    {\kern 0.18em\optbar{\kern -0.18em B}{}\xspace}
\def\DorDbar    {\kern 0.18em\optbar{\kern -0.18em D}{}\xspace}
\def\KorKbar    {\kern 0.18em\optbar{\kern -0.18em K}{}\xspace}

\mathchardef\Upsilon="7107
\def\Y#1S{\ensuremath{\Upsilon{(#1S)}}\xspace}

\def\FourS {\Y4S}

\mathchardef\Deltares="7101
\mathchardef\Xi="7104
\mathchardef\Lambda="7103
\mathchardef\Sigma="7106
\mathchardef\Omega="710A

\def\Deltabar{\kern 0.25em\overline{\kern -0.25em \Deltares}{}\xspace}
\def\Lbar{\kern 0.2em\overline{\kern -0.2em\Lambda\kern 0.05em}\kern-0.05em{}\xspace}
\def\Sigbar{\kern 0.2em\overline{\kern -0.2em \Sigma}{}\xspace}
\def\Xibar{\kern 0.2em\overline{\kern -0.2em \Xi}{}\xspace}
\def\Obar{\kern 0.2em\overline{\kern -0.2em \Omega}{}\xspace}
\def\Nbar{\kern 0.2em\overline{\kern -0.2em N}{}\xspace}
\def\Xb{\kern 0.2em\overline{\kern -0.2em X}{}\xspace}

\def\BR         {{\ensuremath{\cal B}\xspace}}

\def\mes        {\mbox{$m_{\rm ES}$}\xspace}

\def\DeltaE     {\mbox{$\Delta E$}\xspace}

\newcommand{\tev}{\ensuremath{\mathrm{\,Te\kern -0.1em V}}\xspace}
\newcommand{\gev}{\ensuremath{\mathrm{\,Ge\kern -0.1em V}}\xspace}
\newcommand{\mev}{\ensuremath{\mathrm{\,Me\kern -0.1em V}}\xspace}
\newcommand{\kev}{\ensuremath{\mathrm{\,ke\kern -0.1em V}}\xspace}
\newcommand{\ev}{\ensuremath{\mathrm{\,e\kern -0.1em V}}\xspace}
\newcommand{\gevc}{\ensuremath{{\mathrm{\,Ge\kern -0.1em V\!/}c}}\xspace}
\newcommand{\mevc}{\ensuremath{{\mathrm{\,Me\kern -0.1em V\!/}c}}\xspace}
\newcommand{\gevcc}{\ensuremath{{\mathrm{\,Ge\kern -0.1em V\!/}c^2}}\xspace}
\newcommand{\mevcc}{\ensuremath{{\mathrm{\,Me\kern -0.1em V\!/}c^2}}\xspace}

\def\invfb   {\ensuremath{\mbox{\,fb}^{-1}}\xspace}

\def\mus  {\ensuremath{\rm \,\mus}\xspace}

\def\mus        {\ensuremath{\,\mu{\rm s}}\xspace}    

\def\calL{{\ensuremath{\cal L}}\xspace}

\def\to                 {\ensuremath{\rightarrow}\xspace}

\def\pep2{PEP-II}

\def\gsim{{~\raise.15em\hbox{$>$}\kern-.85em
          \lower.35em\hbox{$\sim$}~}\xspace}
\def\lsim{{~\raise.15em\hbox{$<$}\kern-.85em
          \lower.35em\hbox{$\sim$}~}\xspace}

\xspace

\newcommand{\epjBase}        {Eur.\ Phys.\ Jour.\xspace}
\newcommand{\jprlBase}       {Phys.\ Rev.\ Lett.\xspace}
\newcommand{\jprBase}        {Phys.\ Rev.\xspace}
\newcommand{\jplBase}        {Phys.\ Lett.\xspace}

\newcommand{\npBase}         {Nucl.\ Phys.\xspace}
\newcommand{\zpBase}         {Z.\ Phys.\xspace}

\newcommand{\epjc}      [1]  {\epjBase\ C~{\bf #1}}

\newcommand{\npb}       [1]  {\npBase\ B~{\bf #1}}

\newcommand{\plb}       [1]  {\jplBase\ B~{\bf #1}}

\newcommand{\jprl}      [1]  {\jprlBase\ {\bf #1}}
\newcommand{\jprd}      [1]  {\jprBase\ D~{\bf #1}}

\newcommand{\zpc}       [1]  {\zpBase\ C~{\bf #1}}

\def\jetset74   {\mbox{\tt Jetset \hspace{-0.5em}7.\hspace{-0.2em}4}\xspace}

\long\def\inst#1{\par\nobreak\kern 4pt\nobreak
    {\it #1}\par\vskip 10pt plus 3pt minus 3pt}

\begin{document}

{\pagestyle{empty}

\begin{flushright}
\babar-CONF-\BABARPubYear/\BABARConfNumber \\
SLAC-PUB-\SLACPubNumber \\
July 2006 \\
\end{flushright}

\par\vskip 5cm

\begin{center}
\Large \bf \boldmath Measurement of the Branching Fractions for\\ the Decays $B^+\to \rho^+\gamma$, $B^0\to \rho^0\gamma$,
                 and $B^0\to\omega\gamma$
\end{center}
\bigskip

\begin{center}
\large The \babar\ Collaboration\\
\mbox{ }\\
\today
\end{center}
\bigskip \bigskip

\begin{center}
\large \bf Abstract
\end{center}
We present preliminary results of a search for the decays
$B^+\to\rho^+\gamma$, $B^0\to \rho^0\gamma$, and $B^0\to\omega\gamma$.
The analysis is based on data containing \numBB\ \BB\ events
recorded with the \babar\ detector at the PEP-II B factory.
We measure branching fractions of
$\BR(\Bp\to\rhop\gamma) = (\BFrp)\times10^{-6}$
and
$\BR(\Bz\to\rhoz\gamma) = (\BFrz)\times10^{-6}$,
where the first errors are statistical and the second systematic,
and set a 90\% C.L.\ upper limit of
$\BR(\Bz\to\omega\gamma) < \BFomUL\times10^{-6}$.
Assuming isospin relations between the three branching
fractions, these results are used to determine the CKM matrix element
ratio $\VtdVts = \VtdVtsvalLabeled$.

\vfill
\begin{center}

Submitted to the 33$^{\rm rd}$ International Conference
on High-Energy Physics, ICHEP 06,\\
26 July---2 August 2006, Moscow, Russia.

\end{center}

\vspace{1.0cm}
\begin{center}
{\em Stanford Linear Accelerator Center, Stanford University,
Stanford, CA 94309} \\ \vspace{0.1cm}\hrule\vspace{0.1cm}
Work supported in part by Department of Energy contract DE-AC03-76SF00515.
\end{center}

\newpage
} 

\begin{center}
\small

The \babar\ Collaboration,
\bigskip

%
{B.~Aubert,}
{R.~Barate,}
{M.~Bona,}
{D.~Boutigny,}
{F.~Couderc,}
{Y.~Karyotakis,}
{J.~P.~Lees,}
{V.~Poireau,}
{V.~Tisserand,}
{A.~Zghiche}
\inst{Laboratoire de Physique des Particules, IN2P3/CNRS et Universit\'e de Savoie,
 F-74941 Annecy-Le-Vieux, France }
% ...

\end{center}\newpage


\setcounter{footnote}{0}
\section{INTRODUCTION}
\label{sec:Introduction}
\begin{figure}[!h]
\vspace*{8ex}
\begin{center}
\includegraphics{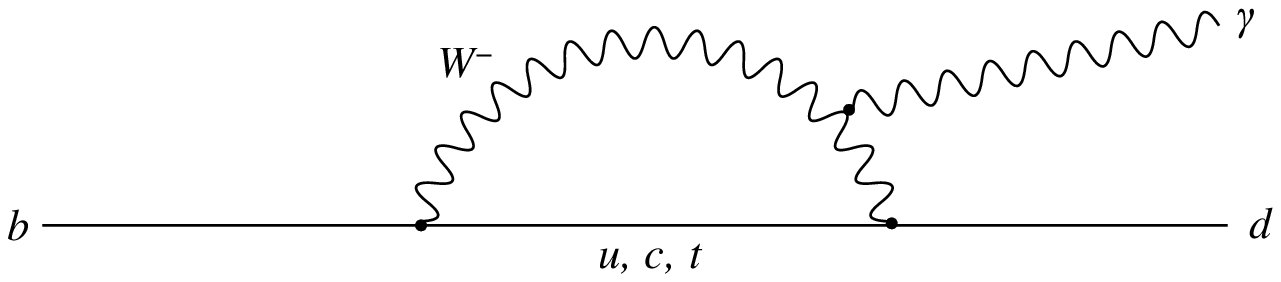}
  \caption{Feynman diagram for $b$ \to $d$\g.}
    \label{fig:feyndiag}
\end{center}
\vspace*{5ex}
\end{figure}

In the Standard Model (SM) description of the decays\footnote{Charge conjugate modes 
are implicitly included throughout.} $\brpg$, $\brzg$,
and $\bomg$, the dominant contributions arise from $\bdg$  penguin
diagrams of the type shown in Figure~\ref{fig:feyndiag}.

\noindent Relating the three individual decay rates by isospin
symmetry and using the measured ratio between the charged and
neutral \B-meson lifetimes $\tau_{\Bp}/\tau_{\Bz}=1.071\pm0.009$~\cite{pdg}, 
one can define a combined branching fraction
\begin{equation}\label{eq:avbr}
    \avbr \equiv 
       \BR(\brpg)
             = 2\frac{\tau_{\Bp}}{\tau_{\Bz}}\BR(\brzg)
             = 2\frac{\tau_{\Bp}}{\tau_{\Bz}}\BR(\bomg)
\end{equation}
The results of recent calculations of $\avbr$ are in the range of
$(0.9$--$1.8)\times 10^{-6}$~\cite{alivtdvtstheory,ali2004,SM}; however, these could
be modified by processes beyond the SM~\cite{hewett}.
Within the SM, 
the isospin violation in these decays is expected to be small; 
a recent estimate~\cite{ali2004} is $(1.1\pm3.9)\%$.

While the exclusive rates have a large uncertainty due to non-perturbative
long-distance QCD effects, much of this uncertainty cancels in the ratio
of $B$ \to $(\rho/\omega)$\g and $B$ \to $K^{*}$\g rates. Since the dominant
diagram in Figure~\ref{fig:feyndiag} involves a virtual top quark, this ratio
is related to the ratio of Cabibbo-Kobayashi-Maskawa (CKM) matrix elements
$\VtdVts$~\cite{alivtdvtstheory,ali2004} via
\begin{equation}\label{eq:vtd}
\frac{\overline{{\cal B}}(B \rightarrow (\rho/\omega)\gamma)}{{\cal B}(B \rightarrow K^{*}\gamma)}= 
\left| \frac{V_{td}}{V_{ts}} \right|^{2}
\left(\frac{1-m_{\rho}^{2}/M_{B}^{2}}{1-m_{K^{*}}^{2}/M_{B}^{2}}\right)^{3}
\zeta^{2} [1+\Delta R].
\end{equation}
Here, the form factor ratio $\zeta$ describes the flavor-SU(3) symmetry
breaking  between $\rho/\omega$ and $K^*$, and $\Delta R$ accounts for
annihilation diagrams.
Physics beyond the Standard Model could affect these decays and create inconsistencies
between the measurements of $\VtdVts$ obtained from this analysis and those obtained
in studies of $B^0_s$ and $B^0_d$ mixing.

Previous searches by $\babar$~\cite{oldbabar} and CLEO \cite{cleo}
have found no evidence for $\B\rightarrow(\rho/\omega)\gamma$ decays. An
observation of the decay $\brzg$ was recently reported by the Belle
collaboration~\cite{newbelle}.

In this paper we report a search for the decays
$\Bp\to\rhop\gamma$, $\Bz\to\rhoz\gamma$, and $\Bz\to\omega\gamma$.
The results presented in this paper use a \babar\ data sample
containing \numBB\ \BB\ events,
corresponding to an integrated luminosity of $316\invfb$,
and supersede those in Ref.~\cite{oldbabar}.

\section{THE \babar\ DETECTOR}
\label{sec:Detector}
The data used in this analysis were collected with the \babar\ detector
at the \pep2\ asymmetric--energy $\epem$ storage ring. Charged particle
trajectories are measured by a combination of a five-layer silicon vertex
tracker and a 40-layer drift chamber in a 1.5-T
magnetic field. Photons and electrons are detected in a CsI(Tl) crystal
electromagnetic calorimeter (EMC) with photon energy resolution
$\sigma_E / E = 0.023 (E/\mathrm{GeV})^{-1/4} \oplus 0.019$. A ring-imaging
Cherenkov detector (DIRC) is used for charged-particle identification.
In order to identify muons, the magnetic flux return is instrumented with
resistive plate chambers and limited streamer tubes. A detailed description
of the detector can be found elsewhere~\cite{ref:babar}.

\section{EVENT RECONSTRUCTION AND SELECTION}
\label{sec:Selection}
The decays \mbox{$\brpg$}, \mbox{$\brzg$}, and \mbox{$\bomg$} are
reconstructed by combining a high-energy photon with a
vector meson exclusively reconstructed in the decays
$\rho^0\to\pip\pim$ ($\BR\approx100\%$),
$\rho^+\to\pip\piz$ ($\BR\approx100\%$),
and $\omega\to\pip\pim\piz$ ($\BR=[89.1\pm0.7]\%$ \cite{pdg}).

The primary source of background is due to continuum events
($\ep\en \to q\bar{q}$, with $q=u,d,s,c$) that contain a
high-energy photon from $\piz$ or $\eta$ decays or from
initial-state radiation (ISR). Decays of $\bkg$, $\Kstar\rightarrow K\pi$
can enter the signal selection, e.g.,  when a $\Kpm$ is misidentified as a $\pipm$.
$\B\rightarrow(\rho/\omega)\piz$ and $\B\rightarrow(\rho/\omega)\eta$
processes are also found to be relevant when a high-energy photon is produced
in the $\piz$ or $\eta$ decay. In addition, there is combinatorial background
from high-multiplicity $\b\rightarrow\s\gamma$ decays.
These backgrounds are suppressed by applying the selection requirements described below.
These requirements have been optimized separately for each signal mode, using simulated
signal and background event samples and the method described in~\cite{SPR},
for maximum statistical sensitivity\footnote{Here, the figure of merit is $S/\sqrt(S+B)$, where $S$ and $B$
are the rates for signal and backgrounds respectively.} assuming a branching fraction  of $1.0(0.5)\times 10^{-6}$
for the charged(neutral) mode.

The photon from the signal $B$ decay is identified as a localized 
energy deposit (cluster) in the calorimeter 
with energy $1.5 <E^*_\gamma <3.5\GeV$ in the  center-of-mass (CM) frame.
The energy deposit must not be
associated with any reconstructed charged track,
be well-isolated from other EMC clusters,
and meet a number of further requirements designed to eliminate background from
hadronic showers, small-angle photon pairs, and charged particles~\cite{babarksg}.

We veto those photons that can be associated with another detected photon to
form a  $\piz$ or $\eta$ candidate using the likelihood ratio
\begin{equation}
\frac{{\cal P}(M(\gamma\gamma'),E_{\gamma'} | i ) }
     {{\cal P}(M(\gamma\gamma'),E_{\gamma'} | \mbox{signal}
       )+{\cal P}(M(\gamma\gamma'),E_{\gamma'} | i ) },\quad i=\piz,\eta .
\end{equation}
In this definition, ${\cal P}$ is the probability density function defined
in terms of the invariant mass of the photon pair, $M(\gamma\gamma')$, and
the energy of $\gamma'$ in the laboratory frame, $E_{\gamma'}$, as determined
from simulated signal and background events.
To consider photons coming from the decays of $\piz$ and $\eta$ that have
converted to $e^+e^-$ pairs,
we combine the high-energy photon candidate with any $\epem$ pair in the
event with an invariant mass $m_{\epem}<50\MeVcc$, and reject the photon
if the total invariant mass satisfies either
$100<m_{\gamma\epem}<160\mevcc$ or $500<m_{\gamma\epem}<590\mevcc$.

Charged pion candidates are selected from well-reconstructed tracks with a
minimum momentum transverse to the beam direction of $100~\mevc$. In order
to reduce backgrounds from charged kaons produced in $\bsg$ processes, a
$\pipm$ selection algorithm \cite{oldbabar} is applied, combining DIRC
information with the energy loss measured in the tracking system.

Photon candidates identified in the EMC with energy greater than $50\mev$
are combined into pairs to form $\piz$ candidates. For $\Bz\rightarrow\omega\gamma$
$(\brpg)$ decays, the invariant mass of the pair is required to satisfy
$122 < m_{\gamma\gamma} < 150\MeVcc$ $(117 < m_{\gamma\gamma} < 148\MeVcc)$. We
also require the cosine of the opening angle between the daughter photons in the
laboratory frame be greater than 0.413 and 0.789 for $\Bz\rightarrow\omega\gamma$
and  $\brpg$ respectively.

The identified pions are combined into vector meson candidates by requiring
$633 < m_{\pip\pim} < 957\MeVcc$, $636 < m_{\pip\piz} < 932\MeVcc$, and
$764 < m_{\pip\pim\piz} < 795\MeVcc$ for $\rho^0$, $\rho^+$, and $\omega$ respectively.
The charged pion pair must originate from a common vertex, which is required to be
consistent with the interaction region to suppress $\KS$ decays.

The photon and $\rho/\omega$ candidates are combined to form the $B$-meson
candidates. We define $\de \equiv E^*_{B}-E_{\rm beam}^*$, where $E^*_B$ is
the CM energy of the $B$-meson candidate and
$E_{\rm beam}^*$ is the CM beam energy.
We also define the beam-energy-substituted mass $\mes \equiv
\sqrt{ E^{*2}_{\rm beam}-{\mathrm{\vec{p}}}_{B}^{\;*2}}$, where ${\mathrm{\vec{p}}}_B^{\;*}$ is
the CM momentum of the $B$ candidate.
Signal events are expected to have a $\de$ distribution centered at zero with a resolution of about
$50\mev$,  and a $\mes$ distribution centered at the mass of the $B$ meson, $m_B$, with a resolution of $3~\mevcc$.
We consider
candidates in the ranges $-0.3 < \de <0.3 \GeV$ and $\mes >  5.22\GeVcc$
to incorporate sidebands that allow the combinatorial background yields
to be extracted from a fit to the data.

To suppress $\B\rightarrow\rho(\piz/\eta)$ and $\B\rightarrow\omega(\piz/\eta)$
events, we calculate the  vector meson helicity angle, $\theta_H$,
defined as the angle between the $\pi^-$ track (normal to the $\omega$ decay plane)
and the B momentum vector in the $\rho$ ($\omega$) rest frame.
We require $|\cos\theta_{H}|< 0.75$.

Contributions from continuum background processes are reduced by considering
only events for which the ratio $R_2$ of second-to-zeroth order Fox-Wolfram
moments~\cite{fox} is less than 0.7. In addition, several variables
that distinguish between signal and continuum events are combined in a neural network.
The quantity $R'_2$, which
is $R_2$ in the frame recoiling against the photon momentum, is used to reject
ISR events. To discriminate between the
jet-like continuum background and the more spherically-symmetric signal
events, we compute the angle between the photon and the thrust axis of the
rest of the event (ROE) in the CM frame. The ROE is defined by all
the charged tracks and neutral energy deposits in the calorimeter that are not
used to reconstruct the $B$ candidate.
We also calculate the moments $L_{i} \equiv \sum_{j} p^{*}_{j}
\cdot|\cos{\theta^{*}_{j}}|^{i}/\sum_{j} p^{*}_{j}$, where $p^{*}_j$ and
$\theta^{*}_{j}$ are the momentum and angle with respect to an axis,
respectively, for each particle $j$ in the ROE. We use $L_{1}$, $L_{2}$,
and $L_{3}$ with respect to the thrust axis of the ROE, as well as
with respect to the photon direction. In addition, we calculate the $B$-meson
production angle $\theta_B^*$ with respect to the beam axis in the CM frame.
Differences in lepton and kaon production
between background and \B decays are exploited by including flavor-tagging  variables~\cite{babartag} as well as the maximum CM momentum and
number  of \Kpm\ and \KS\ in the ROE. The significance of the separation
along the beam axis of the $B$-meson candidate and ROE vertices is included
as well.
To reject events for which this quantity is poorly reconstructed, the separation along the beam axis and
the associated uncertainty are required to be less than 4~mm and 0.4~mm,
respectively.

We train the neural network separately for each signal mode and
select \mbox{$\brpg$}, \mbox{$\brzg$}, and \mbox{$\bomg$} candidates
with a requirement on the the neural-network output that retains
63\%, 74\%, and 71\% of the signal events respectively. For these
cuts, we determine the continuum background efficiencies using a
data sample of $27.2 fb^{-1}$ taken $40\mev$ below the $\FourS$
resonance as $3.0\%$, $5.3\%$ and $6.7\%$ for the three signal modes
respectively.

The expected average candidate multiplicity in the selected signal events is 1.01 for $\brzg$ and
1.07 for $\brpg$ and $\bomg$; in events with multiple candidates
the one with the reconstructed vector meson mass closest to the nominal mass is retained.

Applying all the selection criteria described above, we find efficiencies
of \effrp\% for $\brpg$, \effrz\% for $\brzg$, and \effom\% for $\bomg$.

\section{MAXIMUM LIKELIHOOD FIT}
\label{sec:Fit}
The signal content of the data is determined by means of a multi-dimensional
unbinned maximum likelihood fit, which is constructed individually for each of the
three signal decay modes. All fits use \DeltaE , $\mes$ , $\cos\theta_H$, and
the neural-network output $NN$, after transforming it  according to
\begin{equation}
{\cal NN} = \tanh^{-1} \left( \frac{\left(NN - c_1 \right)
\cdot \left( 1 - c_2 \right) }{c_3} \right),\,\,\,c_i=\mbox{constant}
\end{equation}
in order to facilitate the parameterization of the probability density function (PDF) used in the fit.
For $\bomg$, the cosine of the Dalitz angle $\thd$ \cite{oldbabar} is added
as a fifth observable.

In addition  to signal and continuum background processes, we consider several
sources of background from $B$ decays, which in the fit are combined in
different ways depending on the signal mode under study. In the $\bomg$ fit,
all $B$ backgrounds are combined into a single component, while for the
$\brzg$ analysis $\bkpg$, $\bkgneut$, and other $B$ background processes are
treated separately. The $\brpg$ fit uses four different categories of
$B$ backgrounds:
$\bkpg$ with $K^{* +} \to K^{+} \pi^{0}$, other $\bkg$ decays, $B\to X_s\gamma$
processes (excluding  $\bkg$), and remaining $B$ backgrounds.

In studies of simulated signal and background event samples, the correlations among the
observables are found to be small. We therefore assume that the PDF
$\mathcal{P}(\vec{x_j}; \vec{\alpha_{i}})$ for each of the $N_{\mathrm{hyp}}$
event hypotheses is the product of individual PDFs for the fit observables
$\vec{x}_{j}$  given the set of parameters $\vec{\alpha}_{i}$. The likelihood
function for signal mode $k$ ($=\rhop\gamma$, $\rhoz\gamma$, $\omega\gamma$) is defined as
 \begin{equation}
 {\cal L}_{k}=\exp{\left(-\sum_{i=1}^{N_{\mathrm{hyp}}} n_{i}\right)}\cdot\left
 [\prod_{j = 1}^{N_k}\left(\sum_{i=1}^{N_{\mathrm{hyp}}} n_i{\cal P}_{i}(\vec{x}_j;\ \vec{\alpha}_i)\right)\right]\; ,
 \end{equation}
where $n_i$ is the yield of each hypothesis and $N_k$ is the number of candidate
events observed in data.

The functional form of each PDF is determined from a one-dimensional fit to a 
dedicated sample of simulated
events. The $\de$ distribution is corrected for the observed difference between data
and simulated samples of $\bkg$ decays. All continuum background PDF parameters 
float freely in the fits while the shapes of the signal and $B$ background
distributions are fixed. 
For $\bomg$, the $B$ background yield floats freely in the fit. 
In the $\brpg$ analysis, the  $\bkpg$  ($K^{* +}\to K^{+} \pi^{0}$) contribution and 
the ratio of the other three $B$ background yields are determined from simulated events, 
as are the relative contributions from the three $B$ background components in the 
$\brzg$ fit. 

For the signal, the $\mes$ spectra are described by Crystal Ball functions \cite{CryBall},
the angular distributions are modeled by
second-order polynomials,
and the distributions of $\de$ and ${\cal NN}$ are parametrized as asymmetric,
variable-width Gaussians
\begin{equation}\label{eq:cr}
f(x) = \exp \left[ \frac{-(x-\mu)^2}{2 \sigma^2_{L,R} + \alpha_{L,R} (x-\mu)^2} \right],
\end{equation}
where $\mu$ is the peak position of the distribution, $\sigma_{L,R}$ are the width left and right
of the peak, and $\alpha_{L,R}$ are a measure of the tail on the left and right side of the peak
respectively.

The function (\ref{eq:cr}) also describes the continuum background ${\cal NN}$
shape; the remaining continuum spectra are modeled by ARGUS functions \cite{Argus}
($\mes$) or second- and fourth-order polynomials ($\de$, $\cos\theta_H$, and
$\cos\thd)$. Various functional forms are used to describe the different $B$
background components.

In order to measure the combined branching fraction \avbr, we also
perform a simultaneous fit to the three decay--mode specific data
sets for the effective signal yield $n_{\mathrm{eff}}$, which is
related to signal yields and reconstruction 
efficiencies\footnote{The efficiencies include the daughter
branching fractions.}  obtained from the individual fits via
$n(\brpg)=n_{\mathrm{eff}}\cdot\frac{1}{2}\epsilon(\brpg)$ and
$n(\Bz\rightarrow(\rhoz/\omega)\gamma)=\frac{1}{4}\frac{\tau_{\Bz}}
{\tau_{\Bp}}n_{\mathrm{eff}}\cdot\epsilon(\Bz\rightarrow(\rhoz/\omega)\gamma)$.

\begin{figure}[t]
\includegraphics[width=0.5\linewidth,clip=true]{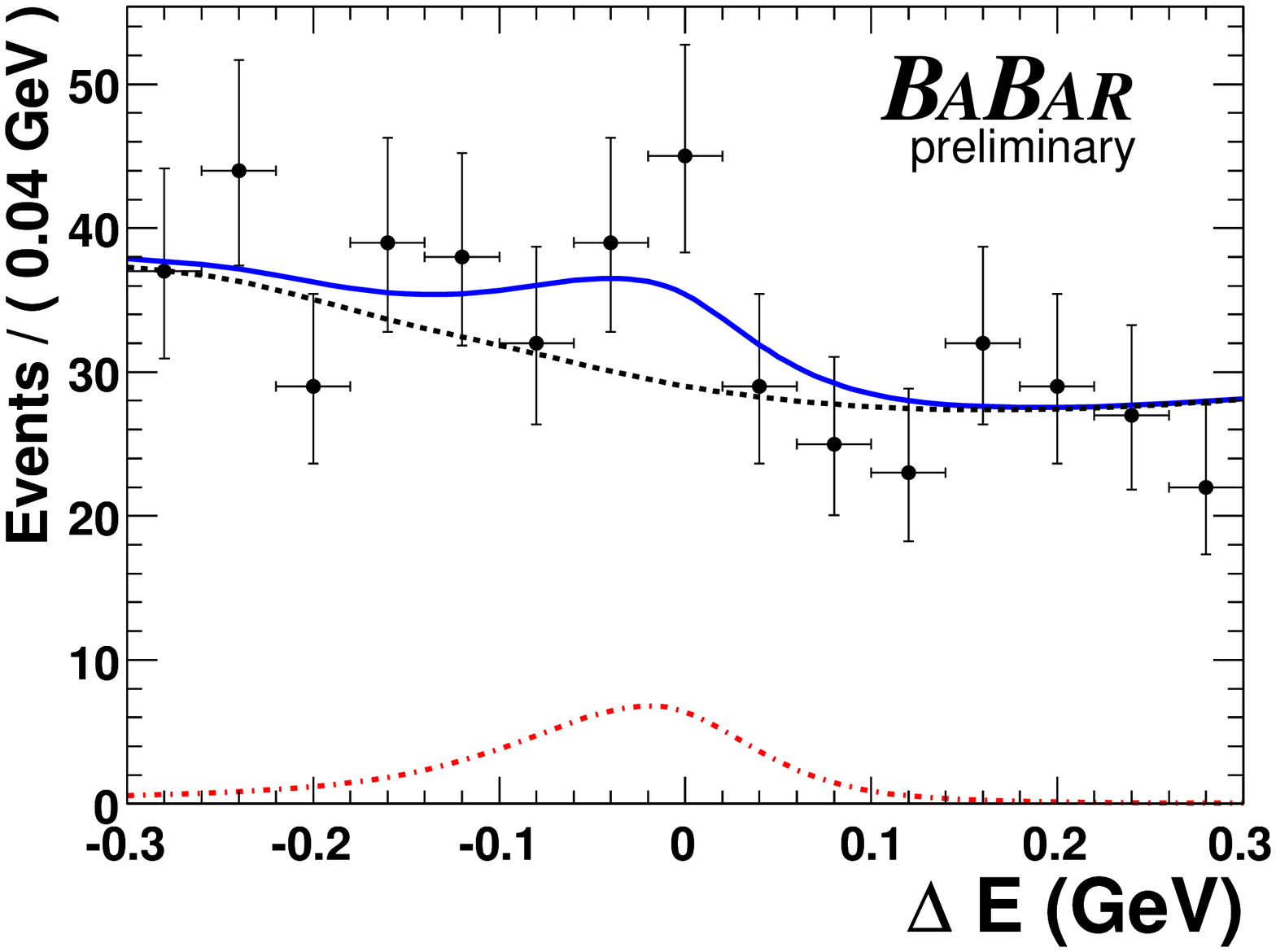}
\includegraphics[width=0.5\linewidth,clip=true]{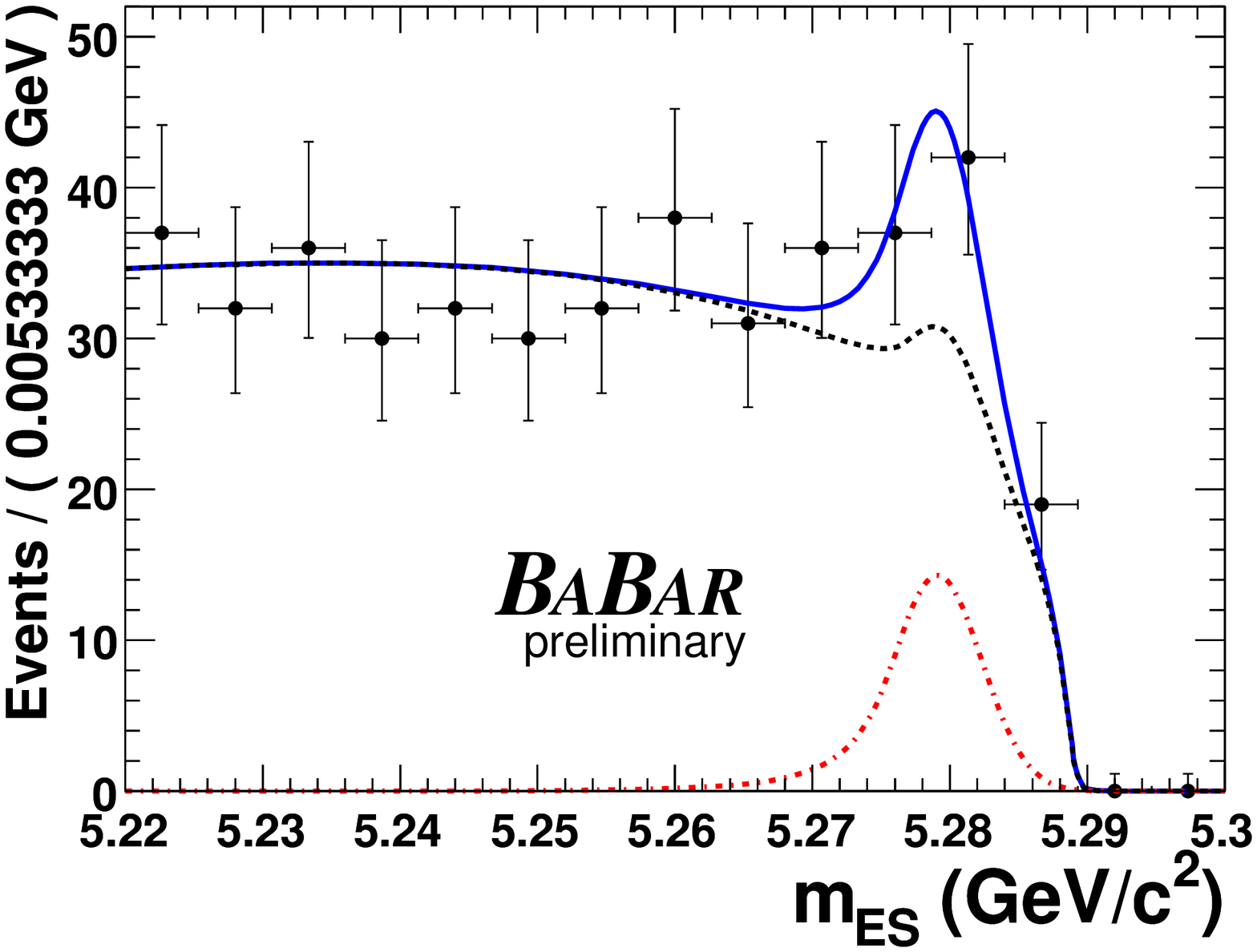}\\
\includegraphics[width=0.5\linewidth,clip=true]{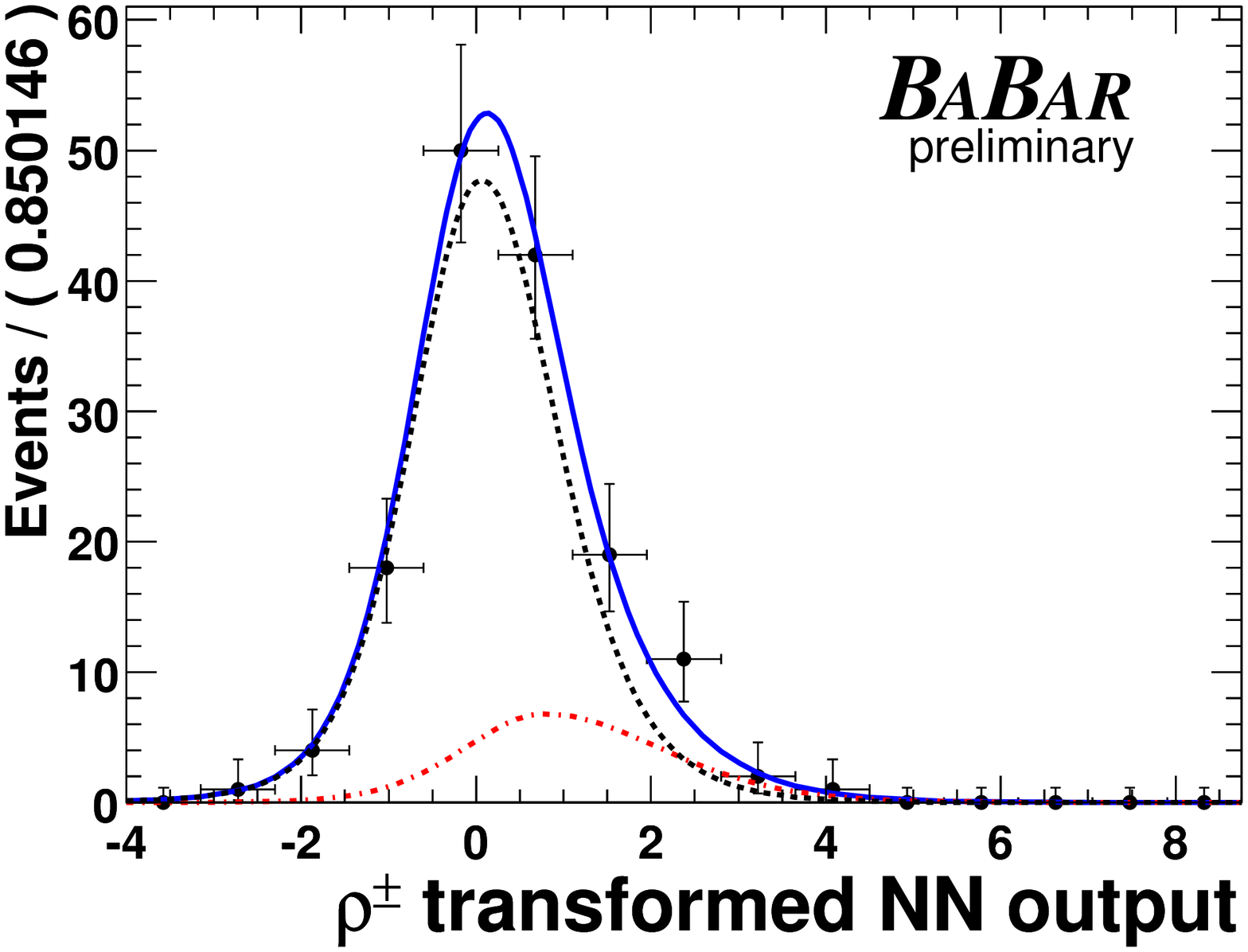}
\includegraphics[width=0.5\linewidth,clip=true]{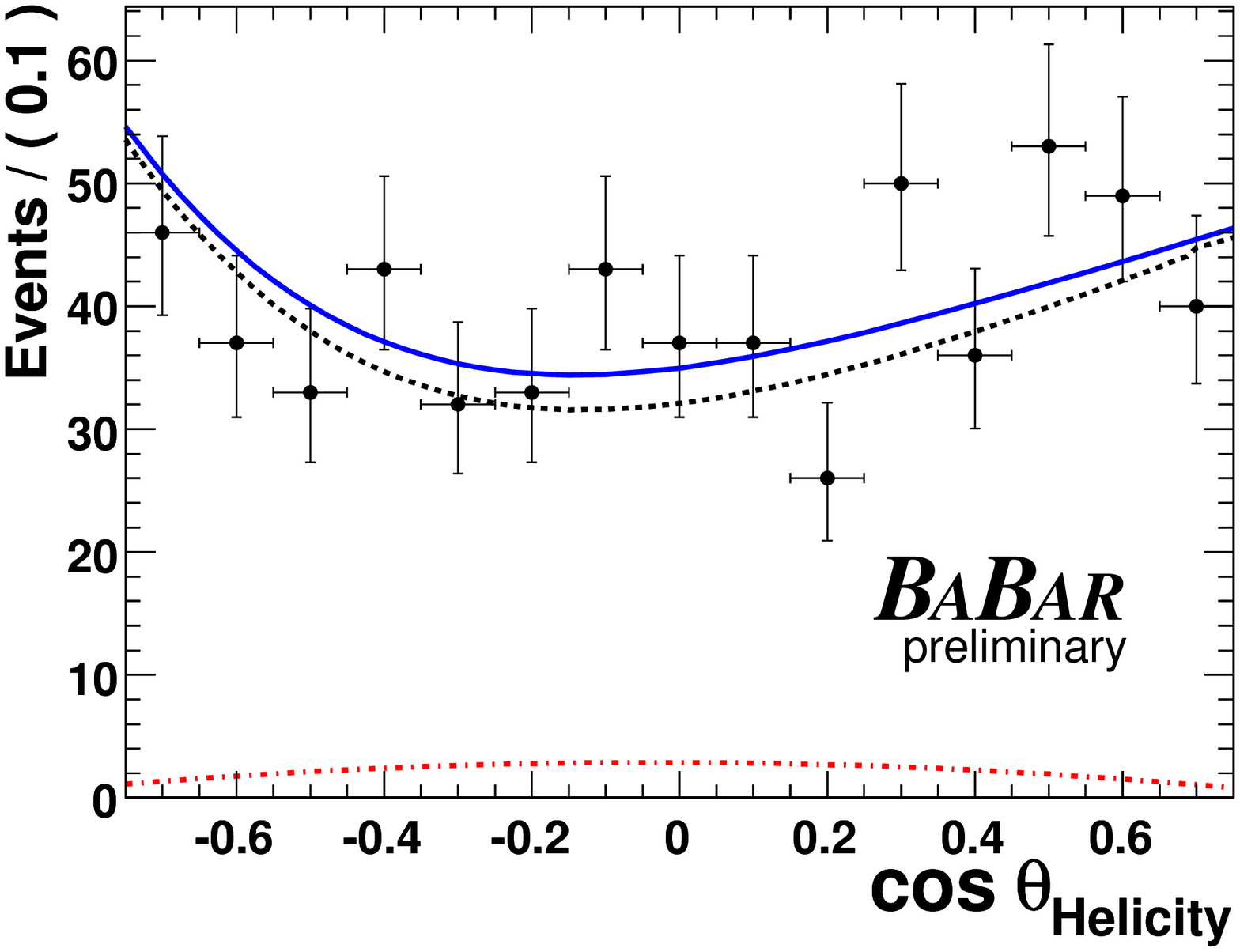}\\
\caption{Projections of the fits to the  $\brpg$  sample in the
discriminating variables $\de$ (upper left), $\mes$ (upper right), $\mathcal{NN}$ (lower left), and
$\cos\theta_H$ (lower right). The points are data, the solid line is the total
PDF and the dark dashed (light dot-dashed) line is the background (signal) only PDF. The selections applied,
unless the variable is projected, are: $-0.15<\de<0.05~\GeV$, $5.275<\mes<5.285~\GeVcc$,
and $\mathcal{NN}>0.0$.
}\label{fig:rhoChfit}
\end{figure}

\begin{figure}[t]
\includegraphics[width=0.5\linewidth,clip=true]{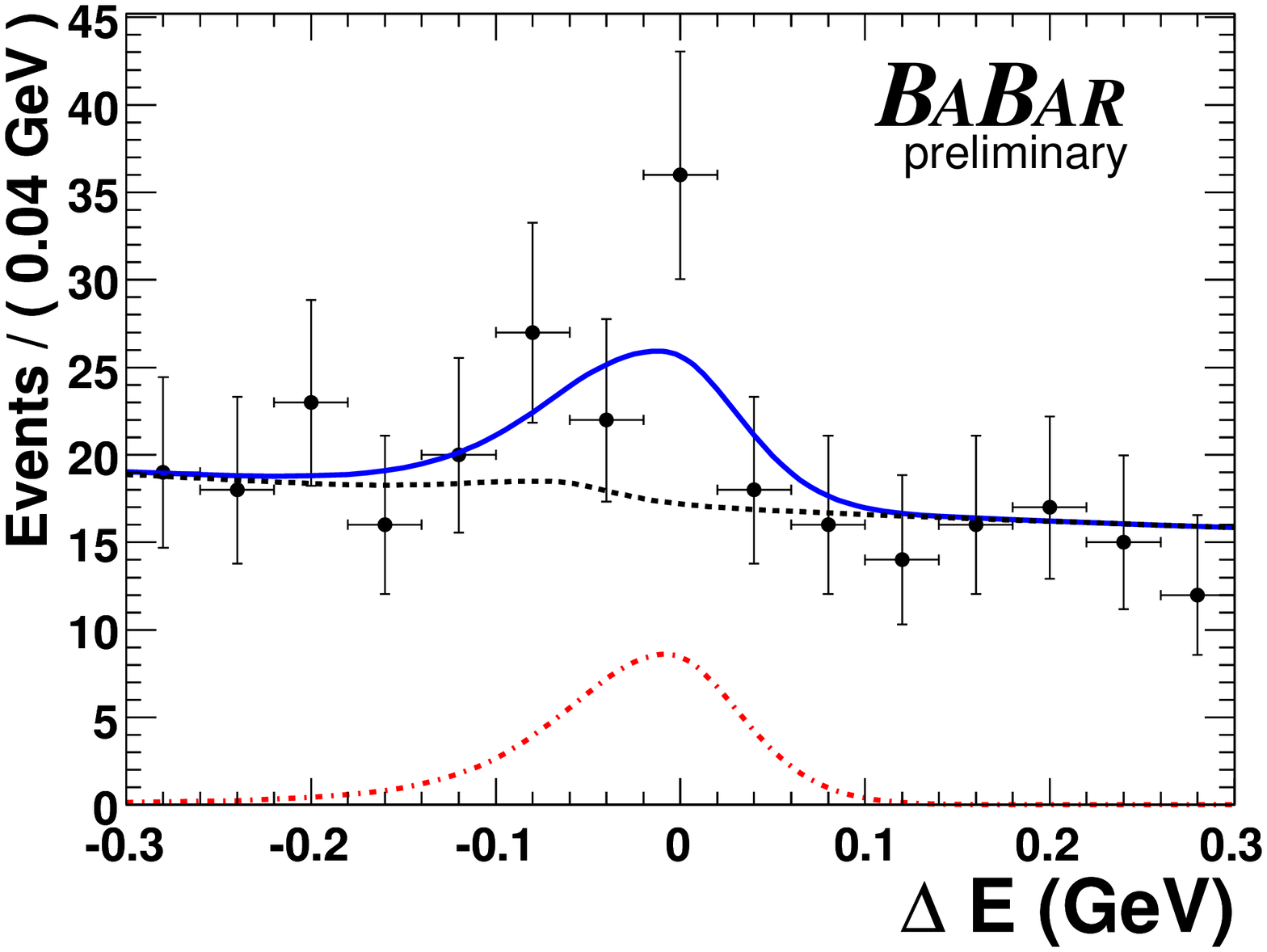}
\includegraphics[width=0.5\linewidth,clip=true]{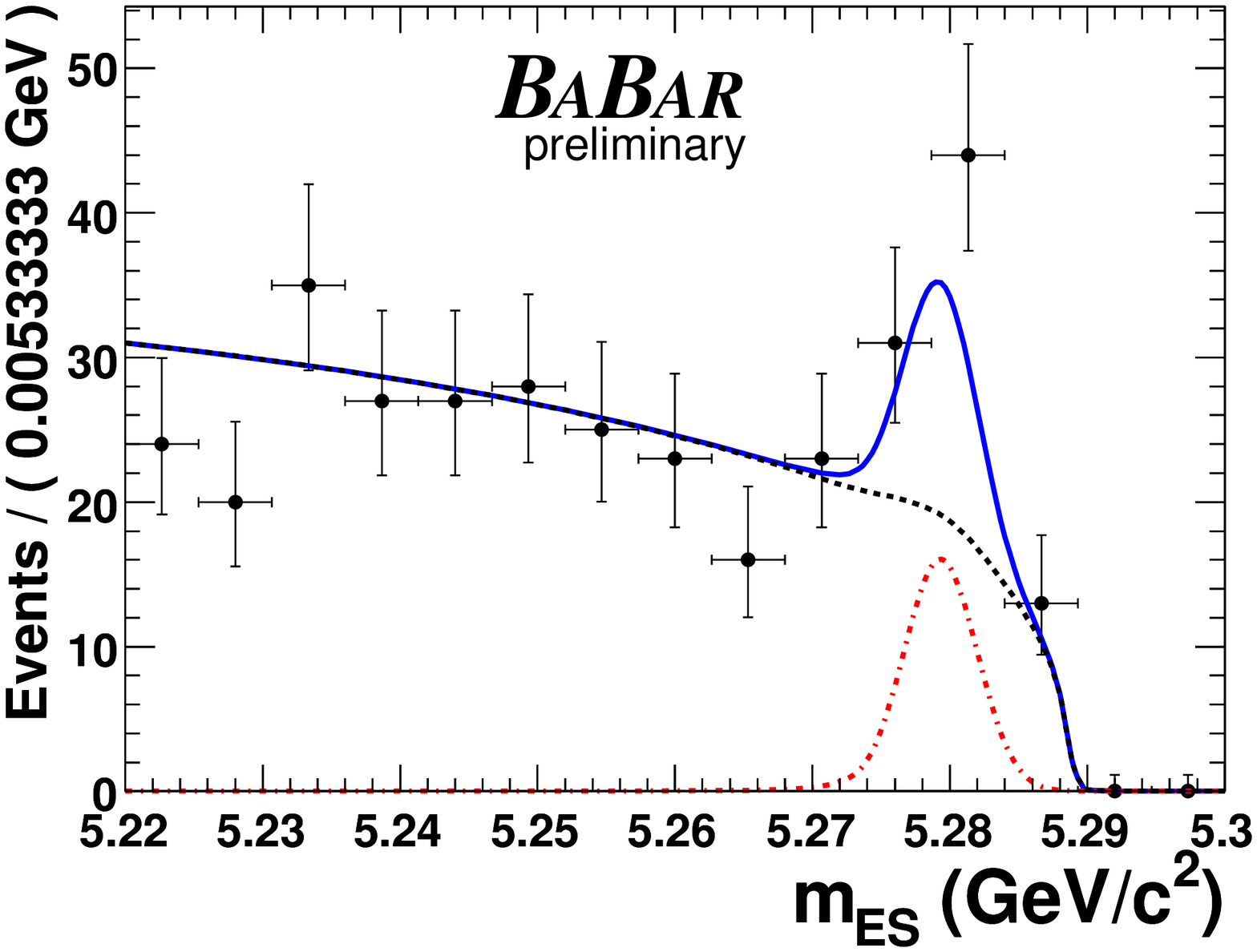}\\
\includegraphics[width=0.5\linewidth,clip=true]{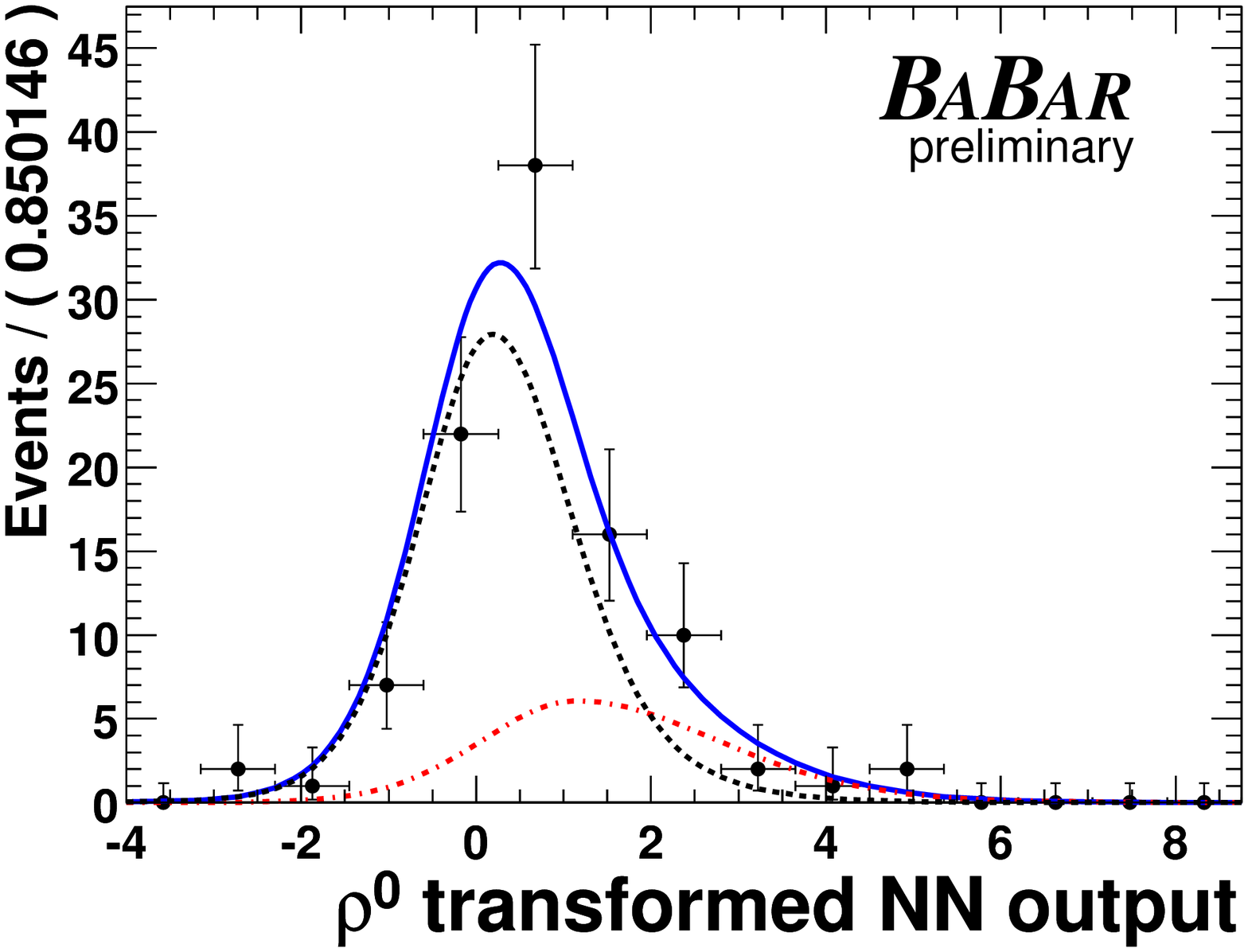}
\includegraphics[width=0.5\linewidth,clip=true]{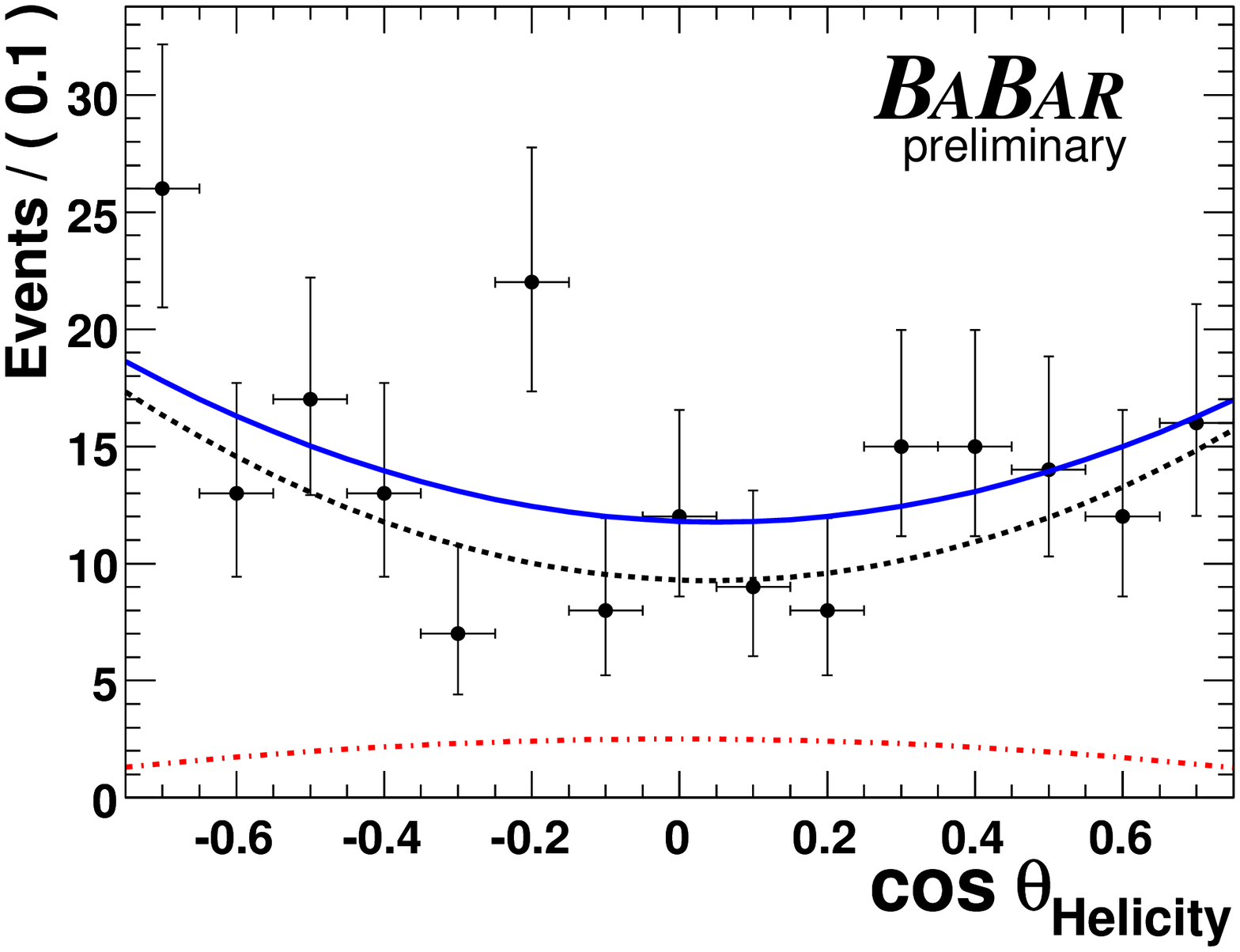}\\
\caption{Projections of the fits to the  $\brzg$  sample in the
discriminating variables $\de$ (upper left), $\mes$ (upper right), $\mathcal{NN}$ (lower left), and
$\cos\theta_H$ (lower right). The points are data, the solid line is the total
PDF and the dark dashed (light dot-dashed) line is the background (signal) only PDF. The selections applied,
unless the variable is projected, are: $-0.15<\de<0.05~\GeV$, $5.275<\mes<5.285~\GeVcc$,
and $\mathcal{NN}>0.0$.
}\label{fig:rho0fit}
\end{figure}

\begin{figure}[t]
\includegraphics[width=0.5\linewidth,clip=true]{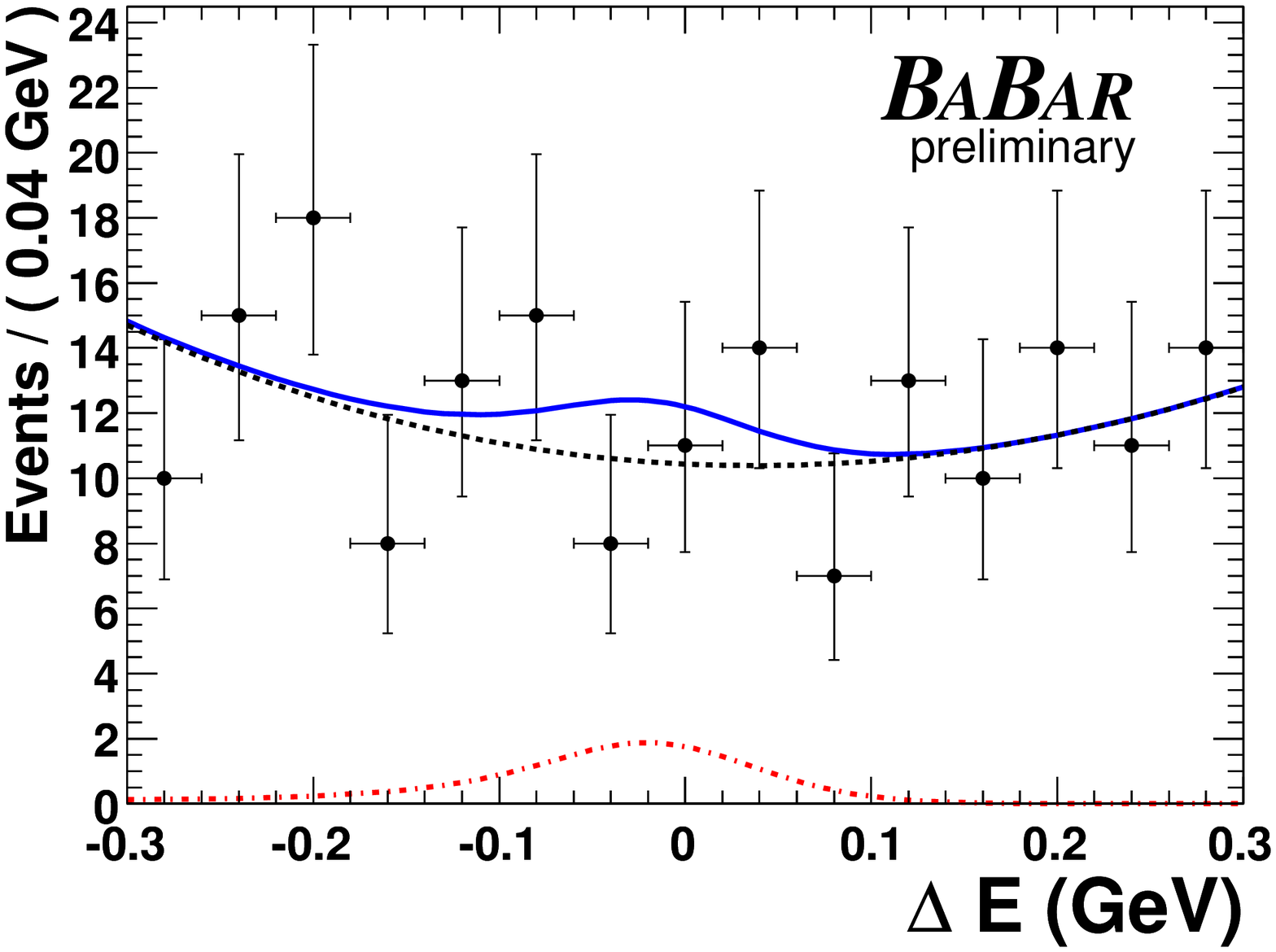}
\includegraphics[width=0.5\linewidth,clip=true]{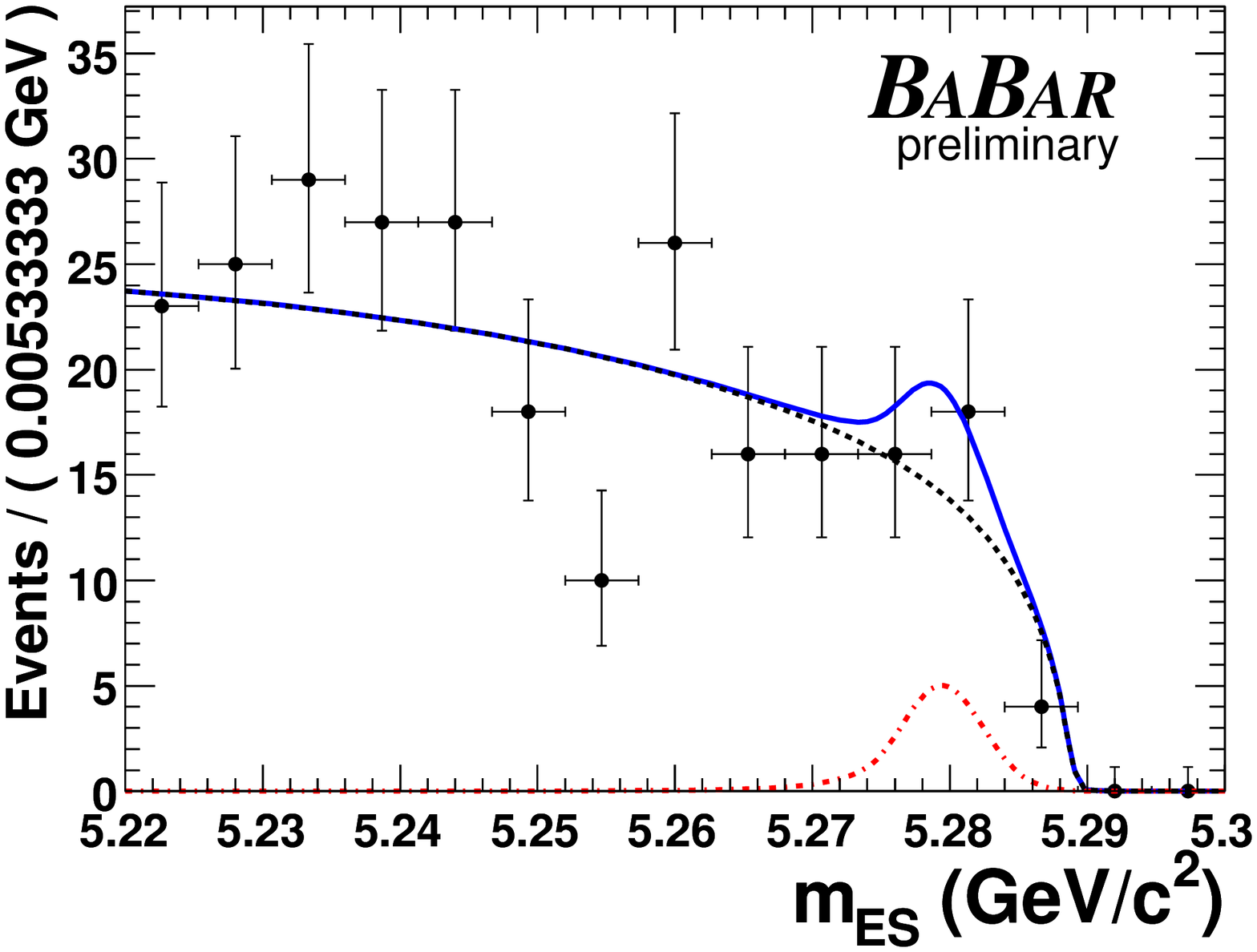}\\
\includegraphics[width=0.5\linewidth,clip=true]{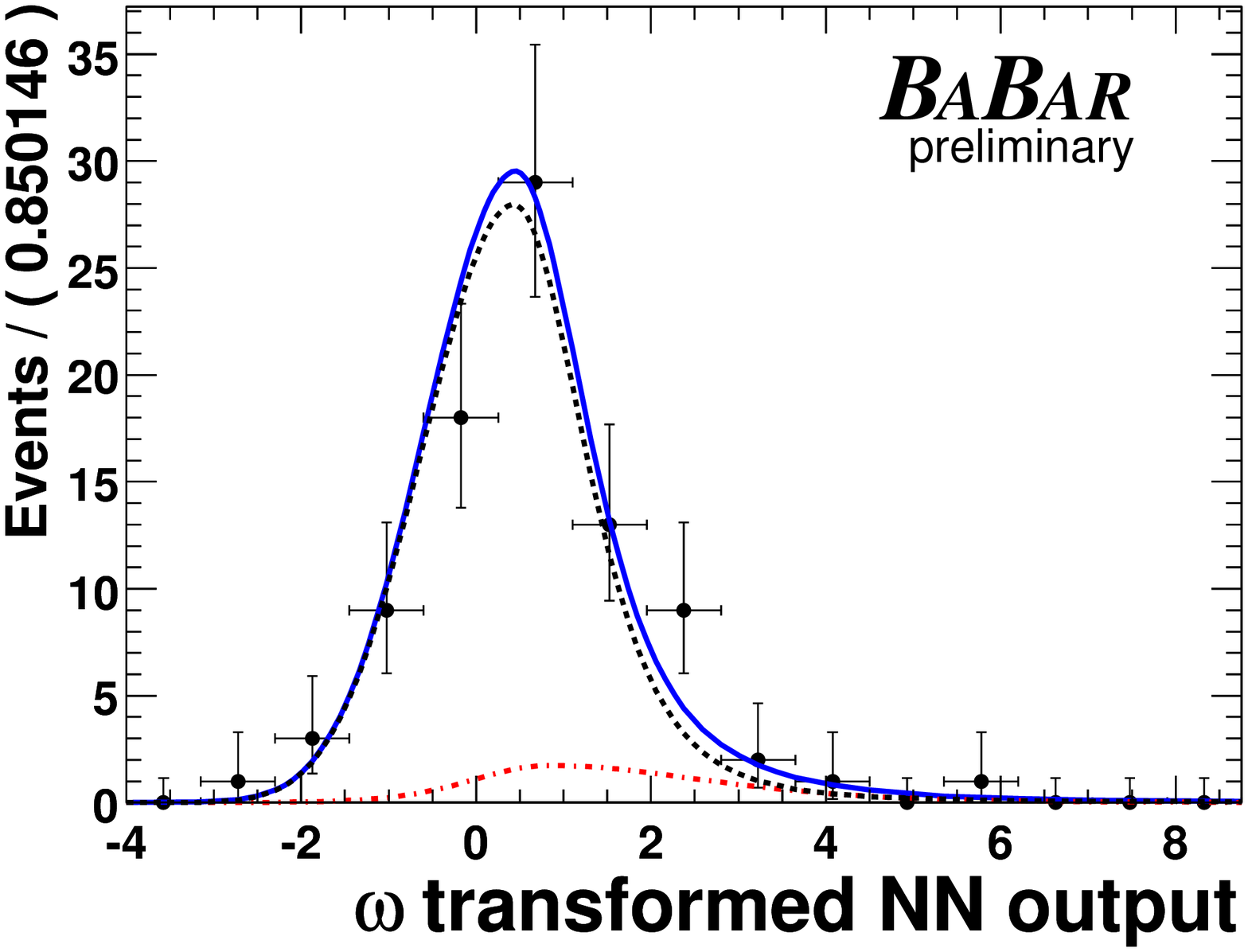}
\includegraphics[width=0.5\linewidth,clip=true]{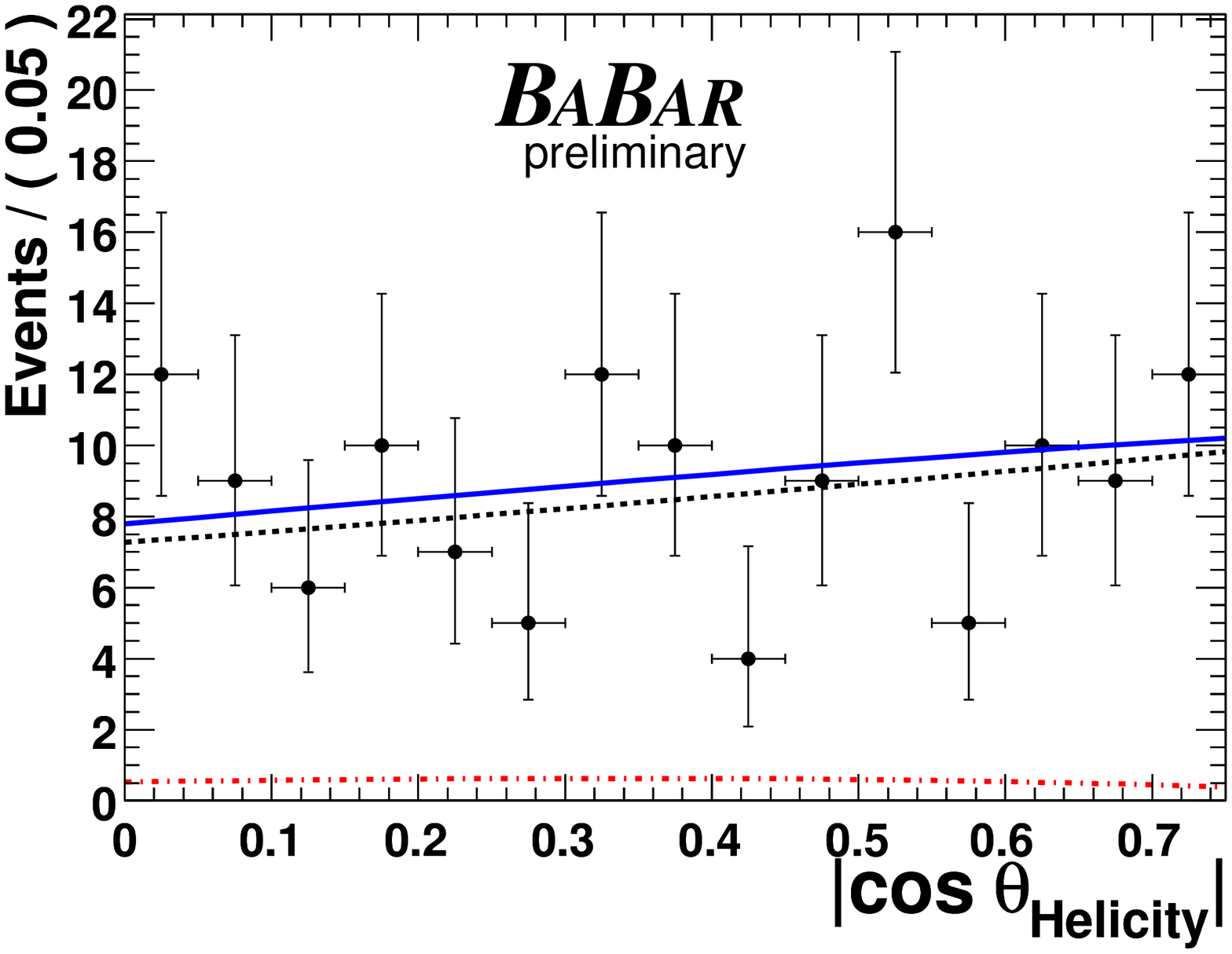}\\
\hspace*{4.2cm}\includegraphics[width=0.5\linewidth,clip=true]{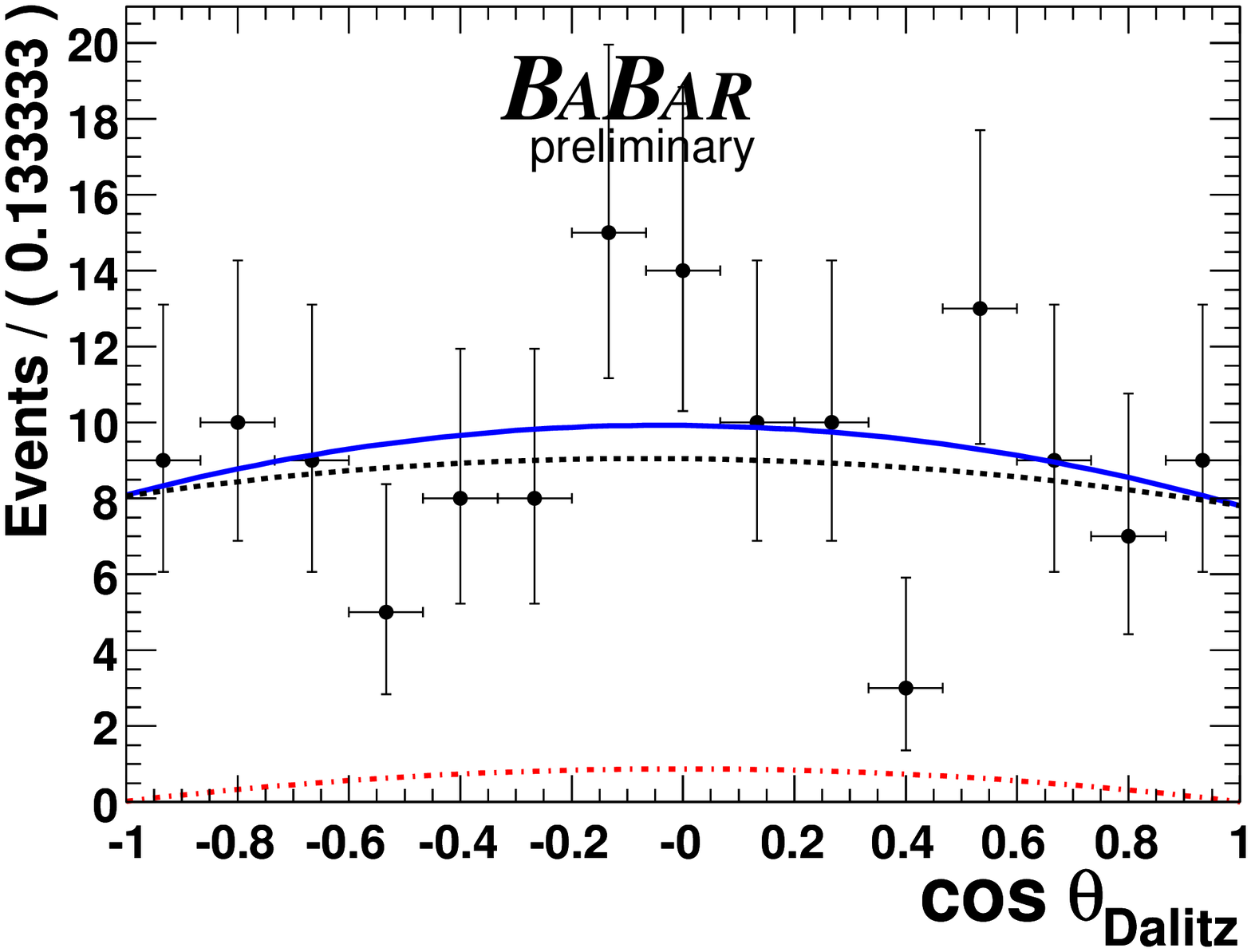}

\caption{Projections of the fits to the  $\bomg$  sample in the
discriminating variables $\de$ (upper left), $\mes$ (upper right), $\mathcal{NN}$ (middle left),
$\cos\theta_H$ (middle right), and  $\cos\theta_D$ (bottom). The points are data, the solid line is the total
PDF and the dark dashed (light dot-dashed) line is the background (signal) only PDF. The selections applied,
unless the variable is projected, are: $-0.15<\de<0.05~\GeV$, $5.275<\mes<5.285~\GeVcc$,
and $\mathcal{NN}>0.0$.
}\label{fig:omegafit}
\end{figure}

Figures \ref{fig:rhoChfit}, \ref{fig:rho0fit}, and \ref{fig:omegafit} show the projections of the fit
results for $\brpg$, $\brzg$, and  $\bomg$ respectively compared to the data; for each plot the signal
fraction is enhanced by selections on the other fit variables.
The resulting signal yields
are given in Table~\ref{tab:results}. The significance is computed as
$\sqrt{2\Delta\log\mathcal{L}}$, where $\Delta\log\mathcal{L}$ is the
log-likelihood difference between the best fit and a fit to the null-signal hypothesis; only statistical 
uncertainties are included here.

\section{SYSTEMATIC UNCERTAINTIES}
\label{sec:Systematics}

Table~\ref{tab:syst} gives an overview of the contributions
to the systematic uncertainties. These are associated with the signal
reconstruction efficiency, the modeling of $\BB$ backgrounds, and the
choice of fixed parameters of the fit PDFs. The latter two contribute
to the uncertainties on the signal yields. A small uncertainty on the 
overall normalization is associated with the imperfect knowledge of the 
total number of $\BB$ pairs in the underlying data sample.
\begin{table}[h]
\vspace*{0ex}
\renewcommand{\arraystretch}{1.3}
\centering
\caption{\label{tab:syst}
Fractional systematic errors (in \%) of the measured branching fractions.}
\vspace*{5ex}
\begin{tabular}{lccc}
\hline
\hline
Source of error & \multicolumn{1}{c}{$ \brpg$}  & \multicolumn{1}{l}{$ \brzg$}
          & \multicolumn{1}{c}{$\bomg$}     \\
\hline
Tracking efficiency     &            1.0\% & 2.0\%        & 2.0\%      \\ 
Charged-particle identification &            2.0\% & 4.0\%        & 2.0\%      \\
Photon selection        &            1.9\% & 2.6\%        & 1.7\%  \\ 
$\pi^0$ reconstruction  &            3.0\% &  -           & 3.0\%      \\
$\pi^0$ and $\eta$ veto &            2.8\% &  $2.8\%$       & 2.8\%      \\
$\mathcal{NN}$ efficiency &                    5.0\% & $3.5\%$      & 3.5\%      \\
$\Delta E$ shape from $K^*\gamma$&         $3.1\%$       & 2.4\%      & 1.9\%      \\
$\mathcal{NN}$ shape      &            $0.2\%$       & 3.9\%      & 4.7\%      \\
$B$ background normalization     &            3.0\% & $4.0\%$          & -      \\
$B$ counting  &            $ 1.1\%$   & 1.1\%     &    1.1\%   \\
\hline
Combined &                            8.4\%      & 9.2\%     & 8.2\%      \\
\hline
\hline
\end{tabular}
\vspace*{2ex}
\end{table}

The signal efficiency systematic error includes uncertainties from tracking,
charged-particle identification, \g/\piz reconstruction, photon selection and
the neural network selection that are determined from suitable independent
data control samples.

To estimate the uncertainty related to the extraction of the signal
PDFs from MC distributions, we vary the parameters within their errors.
The uncertainty related to the choice of a specific functional form for the
shape of the $\mathcal{NN}$ distributions is evaluated by using
a binned histogram as an alternative PDF. All relative and absolute normalizations
of $B$ background components which were fixed in the fit are varied by $50\%$.
For all these variations, the corresponding change in the fitted signal yield is
taken as a systematic uncertainty.

\section{RESULTS}
\label{sec:Results}

The branching fractions are calculated from the fitted signal yields
assuming
$\BR(\Upsilon (4S)\to\BzBzb) = \BR(\Upsilon (4S)\to\BpBm) = 0.5$.
For $\bomg$, we also compute the corresponding $90\%$ confidence level
(C.L.) upper limit using a Bayesian technique. The signal yield upper
limit $n_l$ is determined such that $\int_0^{n_l} {\cal
L}\,dn/\int_0^\infty {\calL}\, dn = 0.90$, assuming a flat prior. The
systematic uncertainty is included by increasing $n_l$ and decreasing
the detection efficiency by their respective errors. The results
are listed in Table~\ref{tab:results}.

\begin{table}
\centering
\caption{\label{tab:results} The signal yield $(n_{\mathrm{sig}})$,
statistical significance in standard deviations ($\sigma$), efficiency $(\epsilon)$,
and branching fraction $(\mathcal{B})$ central value for each mode.
The errors on $(n_{\mathrm{sig}})$ are statistical only, while for the branching fraction
the first errors are statistical and the second systematic.
All results are preliminary.}
\vspace*{5ex}
\renewcommand{\arraystretch}{1.3}
\begin{tabular}{l@{\qquad}ccrcc}
\hline
\hline
Mode & \multicolumn{1}{c}{$n_{\mathrm{sig}}$} & Significance
     & \multicolumn{1}{c}{$\epsilon (\%)$}    & \multicolumn{1}{c}{$\mathcal{\B} (10^{-6})$} \\
\hline
$\brpg$  & $42.4^{+14.1}_{-12.6}$     & $4.1\sigma$ &  $\effrp$ & $\BFrp$ \\
$\brzg$  & $38.7^{+10.6}_{-9.8}$      & $5.2\sigma$ &  $\effrz$ & $\BFrz$ \\
$\bomg$  & $11.0^{+6.7}_{-5.6}$       & $2.3\sigma$ &  $\effom$ & $\BFom$ & ($<\BFomUL$ at 90\% C.L.) \\
\hline
\hline
\end{tabular}
\vspace*{5ex}
\end{table}

The simultaneous fit finds an effective signal yield
$n_{\mathrm{eff}} = 702^{+150}_{-141}$ with a corresponding
statistical significance of 6.3 $\sigma$. This translates into a
combined branching fraction
\begin{equation}\label{eq:res1}
  \avbr= (\BFav)\times10^{-6}.
\end{equation}
We also measure the ratio $\Gamma(\brpg)/[2\Gamma(\brzg)] - 1 =
-0.36 \pm 0.27$ in order to test the hypothesis of isospin symmetry.
The result is in agreement with the theoretical
expectation~\cite{ali2004}.

Using the measured value of \BR(\bkg) \cite{pdg}, we calculate
\begin{equation}\label{eq:res4}
\avbr/\BR(\bkg) = \BrhoBKst.
\end{equation}
This result is used to determine the ratio
of CKM elements $\VtdVts$ by means of Equation~(\ref{eq:vtd}).
Following \cite{Ball:2006nr}, we choose the values $1/\zeta = 1.17 \pm 0.09$,
and $\Delta R = 0.1 \pm 0.1$.
We find
\begin{equation}\label{eq:res5}
\VtdVts = \VtdVtsval,
\end{equation}
where the first error is experimental and the second is theoretical.
This is consistent with the current world average of
$\VtdVts=0.201^{+0.008}_{-0.007}~$\cite{CKMFitter}.

Using the measured value of \BR(\bkog) \cite{pdg}, we  also calculate
\begin{equation}\label{eq:res6}
2 \times \BR(\brzg)/\BR(\bkog) = \BrhozBKstz.
\end{equation}
By only using these two neutral decay modes, the theoretical
interpretation of $\VtdVts$ is simplified since the $W$-annihilation
processes present in the $\brpg$ channel are avoided.
Analogous to Equation~(\ref{eq:vtd}), taking the same values for $1/\zeta$ and $\Delta R$ as above, 
this result is used to obtain
\begin{equation}\label{eq:res7}
\left|V_{td}/V_{ts}\right|_{\rho^0/K^{*0}} = \VtdVtsRzVal,
\end{equation}
where the first error is experimental and the second is theoretical.

\vfill
\section{SUMMARY}
\label{sec:Summary}
In conclusion, we observe the exclusive $\bdg$ transitions
$\brpg$ and $\brzg$ 
and measure the branching fractions 
$\BR(\Bp\to\rhop\gamma) = (\BFrp)\times10^{-6}$
and
$\BR(\Bz\to\rhoz\gamma) = (\BFrz)\times10^{-6}$,
where the first error is statistical and the second is systematic.
We set an improved $90\%$ C.L. upper limit on the $\bomg$ branching fraction of
$\BR(\Bz\to\omega\gamma) < \BFomUL\times10^{-6}$.
Assuming isospin relations between the three branching
fractions, we measure the combined branching fraction $\avbr=
(\BFav)\times10^{-6}$.  
This result translates into a measurement of the CKM matrix element
ratio $\VtdVts = \VtdVtsvalLabeled$. 
In addition, we measure the isospin asymmetry
$\Gamma(\brpg)/[2\Gamma(\brzg)] - 1 = -0.36 \pm 0.27$.
All these preliminary results are consistent within errors with the SM
predictions.

\vfill

\section{ACKNOWLEDGMENTS}
\label{sec:Acknowledgments}
We are grateful for the 
extraordinary contributions of our \pep2\ colleagues in
achieving the excellent luminosity and machine conditions
that have made this work possible.
The success of this project also relies critically on the 
expertise and dedication of the computing organizations that 
support \babar.
The collaborating institutions wish to thank 
SLAC for its support and the kind hospitality extended to them. 
This work is supported by the
US Department of Energy
and National Science Foundation, the
Natural Sciences and Engineering Research Council (Canada),
Institute of High Energy Physics (China), the
Commissariat \`a l'Energie Atomique and
Institut National de Physique Nucl\'eaire et de Physique des Particules
(France), the
Bundesministerium f\"ur Bildung und Forschung and
Deutsche Forschungsgemeinschaft
(Germany), the
Istituto Nazionale di Fisica Nucleare (Italy),
the Foundation for Fundamental Research on Matter (The Netherlands),
the Research Council of Norway, the
Ministry of Science and Technology of the Russian Federation, 
Ministerio de Educaci\'on y Ciencia (Spain), and the
Particle Physics and Astronomy Research Council (United Kingdom). 
Individuals have received support from 
the Marie-Curie IEF program (European Union) and
the A. P. Sloan Foundation.

\vfill
\

\end{document}